\documentclass[cameraready]{Interspeech}

\usepackage{algorithm}
\usepackage{algpseudocode}
\usepackage{multirow}
\usepackage{caption}
\usepackage{subcaption}
\usepackage{booktabs}
\usepackage{amsmath}
\usepackage{graphicx}
\usepackage{flushend}
\usepackage{hyperref}

\title{Spectral Masking and Interpolation Attack (SMIA): A Black-box Adversarial Attack against Voice Authentication and Anti-Spoofing Systems}

\author[affiliation={1}, orcid=0009-0008-5645-4390]{Kamel}{Kamel}
\author[affiliation={1}, orcid=0000-0002-4254-8299]{Hridoy Sankar}{Dutta}
\author[affiliation={1}, orcid=0000-0002-2127-1438]{Keshav}{Sood}
\author[affiliation={1}, orcid=0000-0002-6639-6824]{Sunil}{Aryal}

\address{
    $^1$ School of Information Technology, Deakin University, Australia
}

\email{kamel.kamel@research.deakin.edu.au, hridoy.dutta@deakin.edu.au, keshav.sood@deakin.edu.au, sunil.aryal@deakin.edu.au}


\usepackage{comment}

\begin{document}

\maketitle

\begin{abstract}
    Voice Authentication Systems (VASs) are deeply integrated into high-security sectors, yet remain vulnerable to deepfake and adversarial spoofing attacks. While anti-spoofing countermeasures (CMs) exist to mitigate these risks, many rely on static detection models that can be bypassed by novel adversarial methods, leaving a critical security gap. We propose the \textbf{Spectral Masking and Interpolation Attack (SMIA)}, a black-box adversarial method that strategically manipulates inaudible frequency regions of AI-generated audio to deceive both VAS and CMs simultaneously. Evaluated against state-of-the-art open-source and commercial systems under simulated real-world conditions, SMIA achieves up to 100\% ASR against standalone CMs, 97.5\% against VASs, and 82\% against combined CM and VAS pipelines, conclusively outperforming prior work and underscoring the need for dynamic, adaptive defenses.
\end{abstract}

\section{Introduction}
\label{sec:intro}

Voice Authentication (VA) has emerged as a widely adopted biometric technology due to its ease of use and flexibility in enhancing security within various scenarios including mobile devices, virtual assistants, secure communication systems, financial service institutions, and law enforcement agencies \cite{singh2012applications, algabri2017automatic, rose2006technical}. These VA Systems (VASs) analyze an individual’s voice to create a distinct vocal signature used for identity verification.


Recent advancements in Deep Learning (DL) have revolutionized the VAS, significantly improving accuracy, robustness, and scalability. Modern DL-based approaches have enhanced the field by creating digital voice fingerprints that serve as reference points for authentication, for example, Deep Speaker \cite{deepspeaker} and X-Vectors \cite{xvectors}. Despite these improvements, VAS remain highly vulnerable to various forms of adversarial attacks \cite{lan2022adversarial}.

DL has significantly advanced synthetic speech generation, enabling near-human voice replication through technologies such as Text-To-Speech (TTS) \cite{fish-speech-v1.4, qin2023openvoice, amazon_polly} and Voice Conversion (VC) \cite{li2023freevc, huang2024diffvc+} systems, often using minimal input data. While these methods demonstrate impressive capabilities, they raise serious security concerns when leveraged to deceive or bypass VAS \cite{shonesy2015voice, alali2025partial}. In response, CounterMeasures (CMs) such as RawNet2 \cite{rawnet2}, RawGAT-ST \cite{rawgat}, and RawPC-Darts \cite{rawdarts} have been developed to detect synthetic speech. However, a more sophisticated threat—adversarial spoofing attacks—has emerged, combining high-quality TTS or VC-generated speech with imperceptible adversarial perturbations to simultaneously evade CM detection and bypass the primary VAS, thereby posing a compounded security risk \cite{hai2023sifdetectcracker, kassis2023breaking}.

Adversarial attacks have demonstrated the vulnerability of State-Of-The-Art (SOTA) CMs. For instance, SiFDetectCracker \cite{hai2023sifdetectcracker} achieves over 80\% success by perturbing speaker-irrelevant components, highlighting detectors’ reliance on secondary acoustic cues. However, its impact on full VA stacks—comprising both VAS and CMs—remains unexplored. Similarly, Kassis et al. \cite{kassis2023breaking} proposed a model-agnostic attack targeting the entire VA pipeline via signal-processing transformations, yet its effectiveness drops substantially against hardened defenses (e.g., 15.55\% against AWS Voice ID \cite{amazon2025voiceid} + RawPC-Darts \cite{rawdarts}). These results underscore the persistent challenge of reliably compromising contemporary VA systems.

The primary contributions are summarized as follows:
\begin{itemize}
    \item We introduce the \textbf{Spectral Masking and Interpolation Attack (SMIA)}, a novel black-box adversarial spoofing attack exploits inaudible regions of AI-generated audio to bypass both VAS and CMs.

    \item We conduct evaluations of SMIA against widely adopted SOTA open-source and commercial VASs and CMs.

    \item Our results demonstrate that SMIA significantly outperforms current SOTA attacks by addressing critical methodological gaps, thereby creating a more robust and effective threat.
\end{itemize}



\section{Background and Related Works}
\label{sec:background}

\subsection{Speaker Recognition Systems (SRS)}
SRS identifies individuals through their vocal features. The authentication process typically unfolds in three key phases: (i) \textbf{training phase} - the system learns voice characteristics from a large corpus to develop robust voice authentication models. Next; (ii) \textbf{enrollment phase} - a user's voice is recorded as they speak specific phrases, and the extracted embeddings are stored; and (iii) during the \textbf{verification phase}, the system compares embeddings from the newly captured voice sample to the enrolled embeddings to confirm identity.


\subsection{Anti-Spoofing Countermeasures (CMs)}
Fake voice detection has evolved from handcrafted feature pipelines (MFCCs, CQCCs fed into GMMs/SVMs \cite{patel2015svm,de2017intro}) toward deep learning, with SOTA driven by two paradigms. The first leverages Self-Supervised Learning (SSL) models such as \textit{Wav2Vec2.0} \cite{baevski2020wav2vec} and \textit{WavLM} \cite{chen2022wavlm} as front-ends, paired with lightweight back-end classifiers like LCNN \cite{lavrentyeva2017audio}, or AASIST \cite{jung2022aasist}, though their large-scale pre-training demands significant compute. The second paradigm operates on raw waveforms, exemplified by \textit{RawNet2} \cite{rawnet2}, \textit{RawPC-Darts} \cite{rawdarts}, which applies differentiable NAS to optimize network design, and \textit{RawGAT-ST} \cite{rawgat}, which integrates graph attention with spectro-temporal modeling for robustness against unseen attacks.

\subsection{Adversarial Spoofing Attacks}
These attacks typically involve adding subtle perturbations to AI-generated audio, crafted by speech synthesis models, to fool CMs or both the VAS and CMs.

\textbf{White-box adversarial attacks} have proven highly effective against CM systems when gradient access is available. Liu et al. \cite{liu2019adversarial} showed that minimal FGSM and PGD perturbations can drive CM error rates to 85--95\%, effectively reducing 
detection to random guessing. Building on this, Gomez-Alanis et al. \cite{gomez2022adversarial, gomez2022ganba} extended the threat to joint CM and ASV systems via a trainable perturbation generator (ABTN) and a GAN-based approach, improving attack EER by 27--51\% while preserving speaker identity. However, these methods require internal model access or a surrogate, limiting their applicability.

In the domain of \textbf{black-box adversarial attacks}, Hai et al.\ \cite{hai2023sifdetectcracker} proposed SiFDetectCracker, a framework targeting CMs. Their method generates perturbations for Speaker-irrelevant Features (SiFs), specifically the background noise and the mute parts of an audio sample. Evaluated against detectors including RawNet2\cite{rawnet2}, RawGAT-ST\cite{rawgat}, RawPC-Darts \cite{rawdarts}, and Deep4SNet \cite{ballesteros2021deep4snet}, using the ASVspoof 2019 dataset \cite{Todisco2019ASVspoof}, it achieved an average success rate of 82.2\%. While effective, ablation studies revealed a reliance on both time and noise perturbations simultaneously, and transferability across architectures remained low, a limitation our work aims to address.

While the work by Kassis et al.\ \cite{kassis2023breaking} presents a novel methodology for bypassing VAS, its real-world applicability is limited. Their results against combination systems rely on a cumulative metric insufficiently descriptive of actual effectiveness; performance dropped significantly against SOTA CMs, and the designated-app attack vector assumes a rooted phone, a prerequisite blocked by financial application security.
\section{Methodology}
\label{sec:methodology}

\begin{figure*}[htpb]
    \centering
    \includegraphics[width=0.8\textwidth]{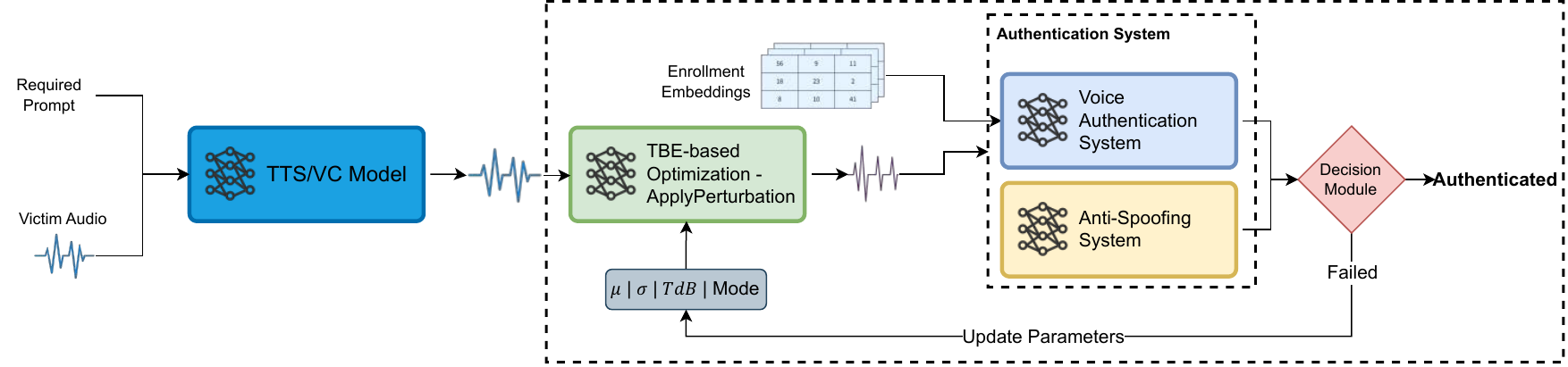}
    \caption{Our Spectral Masking \& Interpolation Attack (SMIA) that targets both Voice Authentication and Anti-Spoofing Systems.}
    \label{fig:workflow}
\end{figure*}

A combined voice authentication and anti-spoofing system processes a given voice sample $x \in \mathbb{R}^m$ to perform two key tasks: \textbf{SV} and \textbf{CM}. The system grants access only when the sample is verified as belonging to the claimed identity \textit{and} determined to be a genuine, live utterance. In this section, we first present \textbf{SMIA}, which manipulates these features to bypass the countermeasure while preserving the speaker characteristics needed to fool the verification model, before delving into its details.




\subsection{Threat Model}
\label{sec:threat_model}
We consider an attacker whose goal is to gain unauthorized access to a victim's SV account by mimicking their voice, operating under a \textbf{black-box assumption} with no internal knowledge of the target VAS or CMs. The attacker collects as little as 10 seconds of victim 
audio from public sources (e.g., social media), generates spoofed audio via any voice-cloning tool (e.g., Fish Speech \cite{fish-speech-v1.4}), and iteratively submits it to the authentication endpoint to optimize against its feedback. We consider two feedback scenarios: \textbf{Score and Label} (a continuous confidence score plus decision, e.g., Azure SV \cite{microsoft_speaker}) and \textbf{Label Only} (a binary pass/fail). The attack is stealthy — a successful login does not alert the legitimate user — and applies to both text-dependent and text-independent SV systems.

\subsection{Overview}

We propose a novel black-box adversarial attack framework designed to circumvent \textbf{VAS} and their \textbf{CMs}. As illustrated in Figure \ref{fig:workflow}, the attack pipeline combines high-fidelity voice synthesis with an adaptive perturbation process consisting of two main stages. First, a base spoofed audio is generated by feeding a victim's voice sample and a target prompt into a tool-independent voice cloning system (e.g., Fish Speech APIs \cite{fish-speech-v1.4}, ASVSpoof tools \cite{Todisco2019ASVspoof}). Second, this synthesized audio is processed by our core mechanism, featuring two components:
\begin{itemize}
    \item \textbf{Black-Box Optimization Framework:} An iterative, feedback-driven loop that queries the target system and uses its predictions to dynamically update the attack parameters.
    \item \textbf{Spectral Masking and Interpolation:} A stealthy perturbation technique that introduces distortions in perceptually insignificant spectral regions to preserve the audio's naturalness.
\end{itemize}

\subsection{Black-Box Optimization (BBO) Framework}
\label{sec:bbo}
This component serves as the ``engine'' of the attack, systematically searching for the optimal parameters for the spectral perturbation module (Section~\ref{sec:apply_perturbations}) without relying on gradient information or internal model access (Section~\ref{sec:threat_model}). To efficiently navigate the search space, we employ a Tree-structured Parzen Estimator (TPE), a Bayesian optimization technique that builds probabilistic models of ``good'' and ``bad'' parameters based on a quantile $J^*$ of observed objective scores:

\begin{align}
    l(\theta) &= p(\theta \mid J(\theta, m) > J^*) \label{eq:tpe_good} \\
    g(\theta) &= p(\theta \mid J(\theta, m) \le J^*) \label{eq:tpe_bad}
\end{align}

By suggesting candidates that maximize the ratio $l(\theta) / g(\theta)$, TPE efficiently focuses the search on the most promising regions. The framework adapts its optimization objective $J(\theta, m)$ based on the feedback tuple $(s, l)$ returned by the target model, where $s$ is the confidence score and $l$ is the decision label:
\begin{equation}
J(\theta, m) = 
\begin{cases} 
    s & \text{if continuous score is available,} \\ 
    1 & \text{if only label } l \text{ is available and } l = \text{success,} \\ 
    0 & \text{if only label } l \text{ is available and } l = \text{fail.} 
\end{cases}
\label{eq:adaptive_objective}
\end{equation}

Crucially, the attack loop terminates immediately once a successful label $l$ is achieved, regardless of the score magnitude. The search space is bounded with the hyperparameter ranges: threshold $t_{db} \in [-60.0, -10.0]$ dB, probability mean $\mu \in [0.0, 1.0]$, and standard deviation $\sigma_{p} \in [0.0, 0.75]$. The optimization runs for a maximum of $N_{iter}=100$ trials (Algorithm~\ref{alg:attacks}).

\begin{algorithm}[h]
\caption{Black-Box Optimization Attack}
\label{alg:attacks}
\begin{algorithmic}[1]
\State \textbf{Input:} Enrolled user ID $U_{id}$, Audio Verification Sample $A_V$, Max iterations $N_{iter}$
\State \textbf{Output:} Attack success status and best result $R_{best}$

\Procedure{Attack}{\textbf{$U_{id}$}, $A_V, N_{iter}$}
    \State $M \gets \{\text{`interpolate', `mask', `hybrid'}\}$ \Comment{Modes}
    \State $S_{best} \gets -\infty$
    \State $R_{best} \gets \text{None}$
    \State $\text{success} \gets \text{False}$
    
    \For{$m \in M$}
        \State Initialize BBO sampler $\mathcal{S}_m$
        \Comment{$\theta = \{t_{db}, \mu, \sigma_p\}$}
        
        \For{$i \gets 1 \text{ to } N_{iter}$}
            \State $\theta \gets \mathcal{S}_m.\text{suggest\_parameters}()$
            \State $A_C \gets \textbf{ApplyPerturbations}(A_V, \theta, m)$
            
            \State $(s, l) \gets \text{Authenticate}(\textbf{$U_{id}$}, A_C)$
            
            \State $\mathcal{S}_m.\text{report\_result}(s, \theta)$
            
            \If{$s > S_{best}$}
                \State $S_{best} \gets s$
                \State $R_{best} \gets (s, l, \theta, m)$
            \EndIf
            
            \If{$l = \text{`success'}$}
                \State $\text{success} \gets \text{True}$
                \State \textbf{break}
            \EndIf
        \EndFor
        
        \If{$\text{success} = \text{True}$}
            \State \textbf{break}
        \EndIf
    \EndFor
    
    \State \Return $R_{best}$
\EndProcedure
\end{algorithmic}
\end{algorithm}

\subsection{Spectral Masking and Interpolation}
\label{sec:apply_perturbations}

At the core of our attack is the \texttt{ApplyPerturbation} module, which introduces subtle, controlled distortions into an audio signal's spectrogram. To ensure the resulting adversarial audio remains natural-sounding and free of obvious artifacts, the module operates exclusively on low-energy (quiet) time-frequency bins that are perceptually less significant to the human ear.

This strategic targeting is governed by three key parameters, $\theta = (t_{\text{db}}, \mu, \sigma_p)$, provided by the optimizer (Algorithm~\ref{alg:apply_perturbations}):
\begin{itemize}
    \item $t_{\text{db}}$: A decibel threshold used to identify the quiet regions of the spectrogram.
    \item $\mu$ and $\sigma_p$: The mean and standard deviation defining a Normal distribution, $\mathcal{N}(\mu, \sigma_p^2)$. For each attack, the perturbation probability $p$ for a quiet bin is sampled from this distribution and clipped to $[0, 1]$.
\end{itemize}

This stochastic sampling of $p$ prevents a fixed, detectable perturbation pattern. To manipulate the selected acoustic features, the module employs three primary techniques: masking, interpolation, and a hybrid approach.

\begin{algorithm}[h]
\caption{ApplyPerturbations: Perturbation Function}
\label{alg:apply_perturbations}
\begin{algorithmic}[1]
\Require Audio signal $x$, parameter vector $\theta$, attack $mode$
\Ensure Adversarial audio $x_{adv}$

\State $\theta = \{ t_{db}, \mu, \sigma_p \}$

\Procedure{SampleProbability}{$\mu, \sigma_p$}
    \State $p_{raw} \gets \text{SampleFromNormal}(\mu, \sigma_p)$
    \State \Return $\text{Clip}(p_{raw}, 0.0, 1.0)$
\EndProcedure

\Procedure{ApplyPerturbations}{$x, \theta, mode$}
    \State $\text{mag}, \text{phase} \gets \text{STFT}(x)$ 
    
    \State $\text{quiet\_mask} \gets \text{mag} < t_{db}$
    
    \State $p \gets \text{SampleProbability}(\mu, \sigma_p)$
    
    \State $\text{random\_mask} \gets \text{UniformRandom}(\text{shape}(\text{mag})) < p$
    \State $\text{final\_mask} \gets \text{quiet\_mask} \land \text{random\_mask}$
    
    \State $\text{mag\_out} \gets \text{mag}$
    
    \If{$mode = \text{`interpolate'}$ or $mode = \text{`hybrid'}$}
        \State $\text{mag\_out} \gets \text{Interpolate}(\text{mag\_out}, \text{final\_mask})$
    \EndIf
    
    \If{$mode = \text{`mask'}$ or $mode = \text{`hybrid'}$}
        \State $\text{mag\_out}[\text{final\_mask}] \gets 0$
    \EndIf
    
    \State $x_{adv} \gets \text{ISTFT}(\text{mag\_out}, \text{phase})$
    
    \State \Return $x_{adv}$
\EndProcedure
\end{algorithmic}
\end{algorithm}

\subsection{Perturbation Modes}

\begin{itemize}
    \item \textbf{Masking:} Silences selected low-energy (quiet) time-frequency bins by resetting their magnitudes to zero, creating subtle distortion through minimal information loss.
    
    \item \textbf{Interpolation:} Replaces targeted quiet bins with values derived from surrounding high-energy ``anchor'' points. It uses 1D linear interpolation ($y(x) = y_0 + (x - x_0) \frac{y_1 - y_0}{x_1 - x_0}$) across time for each frequency channel to reconstruct a temporally smooth and plausible acoustic contour.
    
    \item \textbf{Hybrid:} Applies a mix of both masking and interpolation to the targeted bins, leveraging both techniques for a more complex and varied spectral perturbation.
\end{itemize}

\section{Experimental Setup and Evaluation}
\label{sec:experimental_setup_methodology}

We evaluate the proposed attack using the ASVspoof 2019 LA \cite{Todisco2019ASVspoof} and LibriSpeech \cite{panayotov2015librispeech} datasets. We target three anti-spoofing Countermeasures (CMs)---RawNet2 \cite{rawnet2}, RawGAT-ST \cite{rawgat}, and RawPC-Darts \cite{rawdarts}---and three Voice Authentication Systems (VAS): X-Vectors \cite{xvectors}, DeepSpeaker \cite{deepspeaker}, and the Microsoft Azure Speaker Verification API \cite{microsoft_speaker}. Experiments are conducted on an NVIDIA T4 GPU.

Our evaluation focuses on four key areas:

\begin{enumerate}
    \item \textbf{End-to-End System Evaluation:} Assessment of the combined (CM + VAS) pipeline using ASVspoof (150 trials) and LibriSpeech (40 speakers synthesized via Fish Speech \cite{fish-speech-v1.4}).
    
    \item \textbf{Targeted Component Benchmarking:} Standalone CM performance across 13 attack types (650 utterances) and standalone VAS evaluation across SV, OSI, and CSI tasks.
    
    \item \textbf{Robustness and Efficiency:} Evaluation under over-the-air and over-the-line (VoIP) channel simulations.
    
    \item \textbf{Ablation Studies:} Isolation of core contributions by comparing three perturbation strategies: Masking, Interpolation, and Hybrid.
\end{enumerate}
\section{Results}
\label{sec:results}

\subsection{End-to-End System Evaluation}
Table \ref{tab:end-to-end-asvspoof} shows results on ASVspoof 2019. SMIA achieves up to 97\% ASR against combined VAS/CM systems, substantially outperforming Kassis et al. [19] across all configurations.

\begin{table}[h]
    \centering
    \caption{ASR (\%) against end-to-end systems (CMs+VAS) on the ASVspoof 2019 dataset.}
    \label{tab:end-to-end-asvspoof}
    \resizebox{\columnwidth}{!}{
    \begin{tabular}{lcccc}
        \toprule
        \multirow{2}{*}{\textbf{Group}} & \multirow{2}{*}{\textbf{VAS Model}} & \multicolumn{3}{c}{\textbf{Anti-Spoofing CM}} \\
        \cmidrule(lr){3-5}
         & & \textbf{RawNet2} & \textbf{RawGAT-ST} & \textbf{RawPC-Darts} \\
        \midrule
        \multirow{2}{*}{Ours} & DeepSpeaker  & 59 & 71 & 30 \\
        & X-Vectors    & \textbf{97} & \textbf{93} & \textbf{82} \\
        \midrule
        \multirow{1}{*}{Kassis et al.} & X-Vectors & - & 14.25 & 9.67 \\
        \bottomrule
    \end{tabular}}
\end{table}

Table \ref{tab:end-to-end-libri} shows results on LibriSpeech, where SMIA achieves 100\% ASR in the majority of configurations. The notable drop to 66.5\% for DeepSpeaker + RawPC-Darts reflects a fundamental tension: perturbations strong enough to fool the robust RawPC-Darts CM degrade the speaker biometric, causing VAS verification to fail.

\begin{table}[h]
    \centering
    \caption{ASR (\%) against end-to-end systems (CMs+VAS) on the LibriSpeech dataset.}
    \label{tab:end-to-end-libri}
    \begin{tabular}{lccc}
        \toprule
        \multirow{2}{*}{\textbf{VAS Model}} & \multicolumn{3}{c}{\textbf{Anti-Spoofing CM}} \\
        \cmidrule(lr){2-4}
         & \textbf{RawNet2} & \textbf{RawGAT-ST} & \textbf{RawPC-Darts} \\
        \midrule
        X-Vectors    & 100 & 100 & 92.5 \\
        DeepSpeaker  & 93.5 & 93.5 & 66.5 \\
        \bottomrule
    \end{tabular}
\end{table}

\subsection{Component-Level Evaluation}
Table \ref{tab:targeted_asvspoof} benchmarks SMIA against CMs on ASVspoof 2019. SMIA substantially outperforms both baselines: 96.3\% vs. 80.4\% on RawNet2 and 95.0\% vs. 75.8\% on RawGAT-ST. Against the harder RawPC-Darts, SMIA achieves 87.0\%, matching SiFDetectCracker (84.1\%) while outperforming Kassis et al. On LibriSpeech, SMIA achieves 100\% ASR against RawNet2 and RawGAT-ST, and 87.9\% against RawPC-Darts.

\begin{table}[h]
    \centering
    \caption{Performance comparison (ASR) of SMIA against SOTA attacks on the ASVspoof 2019 dataset.}
    \label{tab:targeted_asvspoof}
    \resizebox{\columnwidth}{!}{
    \begin{tabular}{lccc}
        \toprule
        \multirow{2}{*}{\textbf{Attack}} & \multicolumn{3}{c}{\textbf{Anti-Spoofing CM}} \\
        \cmidrule(lr){2-4}
         & \textbf{RawNet2} & \textbf{RawGAT-ST} & \textbf{RawPC-Darts} \\
        \midrule
        \textbf{SMIA}    & \textbf{96.3} & \textbf{95} & \textbf{87} \\
        SiFDetectCracker \cite{hai2023sifdetectcracker}    & 80.4 & 75.8 &  84.1 \\
        Kassis et al. \cite{kassis2023breaking}  & -- & 12.03 & 15.82 \\
        \bottomrule
    \end{tabular}}
\end{table}

Table \ref{tab:tasks} presents VAS-only results across SV, CSI, and OSI tasks. SMIA achieves at least 87\% ASR across all models and tasks, including 98\% against the commercial Microsoft Azure SV API, demonstrating broad real-world applicability.

\begin{table}[h]
    \centering
    \caption{Performance across SV, CSI, and OSI tasks.}
    \label{tab:tasks}
    \small
    \resizebox{\columnwidth}{!}{
    \begin{tabular}{llccccc}
        \toprule
        \textbf{Attack} & \textbf{Task} & \textbf{Model} & \textbf{Recall} & \textbf{ASR} & \textbf{F1-score} & \textbf{EER} \\
        \midrule
        \multirow{6}{*}{\textbf{SMIA}} & SV & Deep Speaker  & 0.99 & 97.5 & 0.67 & 0.49 \\
         & SV & X-Vectors  & 1 & 100  & 0.67 & 0.5 \\
         & SV & Microsoft Azure & 0.76 & 98 & 0.56 & 0.6 \\
         & CSI & Deep Speaker  & 0.96 & 87 & 0.67 & 0.45 \\
         & CSI & X-Vectors  & 0.98 & 97   & 0.66 & 0.49 \\
         & OSI & Deep Speaker  & 0.95 & 88 & 0.68 & 0.46 \\
         & OSI & X-Vectors  & 0.98 & 97   & 0.66 & 0.49 \\
        \midrule
        \textbf{Kassis et al. \cite{kassis2023breaking}} & SV & X-Vectors & - & 100 & - & - \\
        \bottomrule
    \end{tabular}}
\end{table}

\subsection{Robustness in Real-World Scenarios}
Under simulated over-the-air and over-the-line transmission, SMIA remains highly effective: 100\% ASR for X-Vectors and 99.5\% for DeepSpeaker over-the-line, and 100\% for both models over-the-air. Slight degradation confirms applicability.

\begin{table}[h]
    \centering
    \caption{Robustness of the SMIA attack under simulated over-the-line and over-the-air transmission channels.}
    \label{tab:practicability}
    \small
    \resizebox{\columnwidth}{!}{
    \begin{tabular}{lccccc}
        \toprule
        \textbf{Channel} & \textbf{Model} & \textbf{Recall} & \textbf{ASR(\%)} & \textbf{F1-score} & \textbf{EER} \\
        \midrule
        Over-the-line & Deep Speaker & 0.99 & 99.5 & 0.66 & 0.49 \\
        Over-the-line & X-Vectors & 1 & 100 & 0.67 & 0.5 \\
        Over-the-air & Deep Speaker & 1 & 100 & 0.67 & 0.5 \\
        Over-the-air & X-Vectors & 1 & 100 & 0.67 & 0.5 \\
        \bottomrule
    \end{tabular}}
\end{table}



\subsection{Ablation Study on Attack Modes}
Table \ref{tab:ablation_modes} compares the three perturbation modes. Interpolation dominates on RawNet2 and RawGAT-ST (100\% ASR with X-Vectors) but drops to 5\% against RawPC-Darts. Masking yields moderate, consistent performance. Hybrid achieves 84.4\% ASR against RawPC-Darts, making it the most robust overall. These findings motivate the multi-mode search in Algorithm \ref{alg:attacks}.

\begin{table}[htpb]
    \centering
    \caption{Ablation study of SMIA attack modes. The table shows the ASR (\%) for the Masking, Interpolation, and Hybrid modes against various combined CM and VAS systems.}
    \label{tab:ablation_modes}
    \resizebox{\columnwidth}{!}{
    \begin{tabular}{llccc}
        \toprule
        \multirow{2}{*}{\textbf{Attack Mode}} & \multirow{2}{*}{\textbf{VAS Model}} & \multicolumn{3}{c}{\textbf{Anti-Spoofing CM}} \\
        \cmidrule(lr){3-5}
         & & \textbf{RawNet2} & \textbf{RawGAT-ST} & \textbf{RawPC-Darts} \\
        \midrule
        \multirow{2}{*}{\textbf{Masking}}  & X-Vectors      & 87 & 75 & 78 \\
                                          & DeepSpeaker    & 76 & 72 & 54 \\
        \midrule
        \multirow{2}{*}{\textbf{Interpolation}}  & X-Vectors      & 100 & 100 & 5 \\
                                          & DeepSpeaker    & 97.7 & 88 & 2 \\
        \midrule
        \multirow{2}{*}{\textbf{Hybrid}}  & X-Vectors      & 94.4 & 71.1 & 84.4 \\
                                          & DeepSpeaker    & 73.3 & 72.2 & 51.1 \\
        \bottomrule
    \end{tabular}}
\end{table}


\section{Discussion and Conclusion}
SMIA exposes the vulnerabilities of current voice security defenses to adaptive black-box attacks. The core insight is that targeting perceptually inaudible spectral regions allows SMIA to simultaneously evade anti-spoofing detection and preserve biometric identity. Furthermore, its ability to generalize across 13 diverse ASVspoof 2019 attack types shows it does not rely on the artifacts of any single synthesis method. \textbf{Future work} should focus on improving perturbation efficiency and using SMIA-generated examples to adversarially train next-generation detection models. \textbf{In conclusion}, SMIA achieves 100\% ASR against standalone CMs, 97.5\% against VAS, and 82\% against combined pipelines. These findings constitute an urgent call for a paradigm shift away from static detection toward dynamic security frameworks capable of co-evolving with the adversarial threat landscape.

\section{Generative AI Use Disclosure}
During the preparation of this work, Generative AI tools (Large Language Models) were used solely for editing and polishing purposes, and not for writing any major part of the manuscript. All outputs were carefully inspected by the authors to ensure accuracy and originality.

\bibliographystyle{IEEEtran}
\bibliography{mybib}

\end{document}